\documentclass{moriond}

\usepackage{mciteplus}
\usepackage{ifthen}
\newboolean{articletitles}
\setboolean{articletitles}{false} 
\def\be{\begin{equation}}
\def\ee{\end{equation}}
\def\bea{\begin{eqnarray}}
\def\eea{\end{eqnarray}}

\newcommand{\Photo}{\includegraphics[height=35mm]{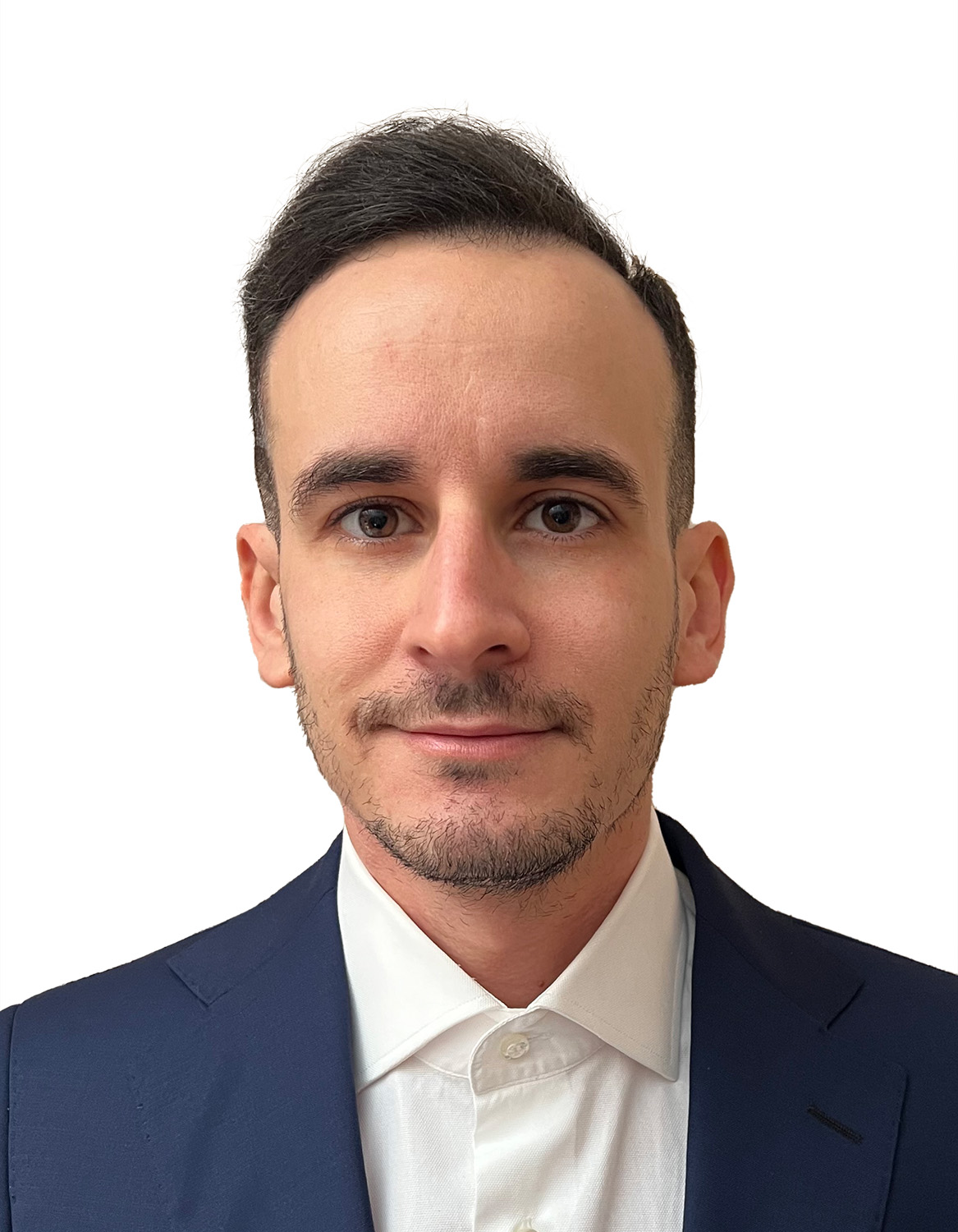}}

\begin{document}
\vspace*{4cm}
\title{Heavy-ion and fixed-target physics at LHCb}

\author{ Andrea Merli$^{1,2}$ }

\address{$^1$Universit\`a degli Studi di Milano, Milano, Italy\\
$^2$ INFN Sezione di Milano, Milano, Italy}

\maketitle\abstracts{
The LHCb collaboration pursues a full physics program studying dense QCD with both beam-beam and fixed-target collisions. The forward design of the LHCb spectrometer allows probing the low-x region of the nucleus, while high vertexing precision and full particle ID guarantee the reconstruction of a wide range of hadrons down to very low transverse momentum. In this contribution we present the recent LHCb results including open charm and charmonia production in fixed target collisions and Bose-Einstein correlation effects in proton-lead collisions.}

\section{Introduction}

The LHCb experiment~\cite{Alves:1129809} has a unique heavy-ion and fixed-target physics program. An extended heavy-ion collision programme in Runs 3 and 4 of the LHC is motivated by the lack of theoretically clean and precise measurements at low $Q^2$  for Bjorken-x down to $10^{-5}$, which provide information about the initial stages of collisions involving heavy ions. In particular, nuclear parton distribution functions (nPDFs) are not well known in the low-x regime. Measurements at forward rapidity at the highest available collision energies are uniquely capable of exploring the low Bjorken-x region. The LHCb experiment offers the unique opportunity to study heavy-ion interactions in the forward region ($2 <\eta< 5$), in a kinematic domain complementary to the other 3 large experiments at the LHC. The detector has excellent capabilities for reconstructing quarkonia and open charm states down to zero $p_T$. It can separate the prompt and displaced charm components. In $p$Pb collisions, both forward and backward rapidities are covered thanks to the possibility of beam reversal. Since 2015, the LHCb forward spectrometer has the unique capability to function as a fixed-target experiment (SMOG, System for Measuring Overlap with Gas) by injecting gas into the LHC beam pipe while proton or ion beams are circulating. LHCb has collected a wide variety of heavy-ion and fixed-target data in recent years. In the standard collision mode, lead-lead (PbPb) and proton-lead ($p$Pb) data have been acquired at $\sqrt{s_{NN}} = 5.02\textnormal{-}8.16 \;\mathrm{TeV}$ with centrality up to $\sim60\%$ for PbPb collisions, limited by the hardware saturation due to the high track density in the forward region. In the fixed-target configuration, several samples of beam-gas collisions were acquired from 2015 to 2018 at $\sqrt{s_{NN}} = 68.5\textnormal{-}110 \;\mathrm{GeV}$, acting as the highest-energy fixed-target experiment ever.



\section{Bose-Einstein
correlations of identical pions in
proton-lead collisions}

\begin{figure}[htb]
    \centering
    \includegraphics[width=0.4\textwidth]{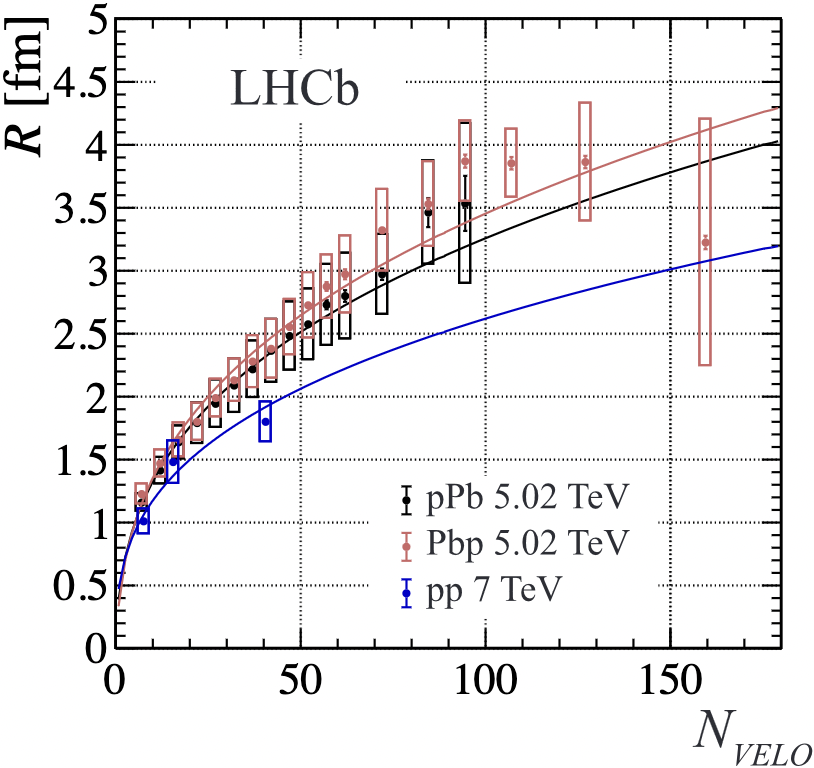}        \includegraphics[width=0.4\textwidth]{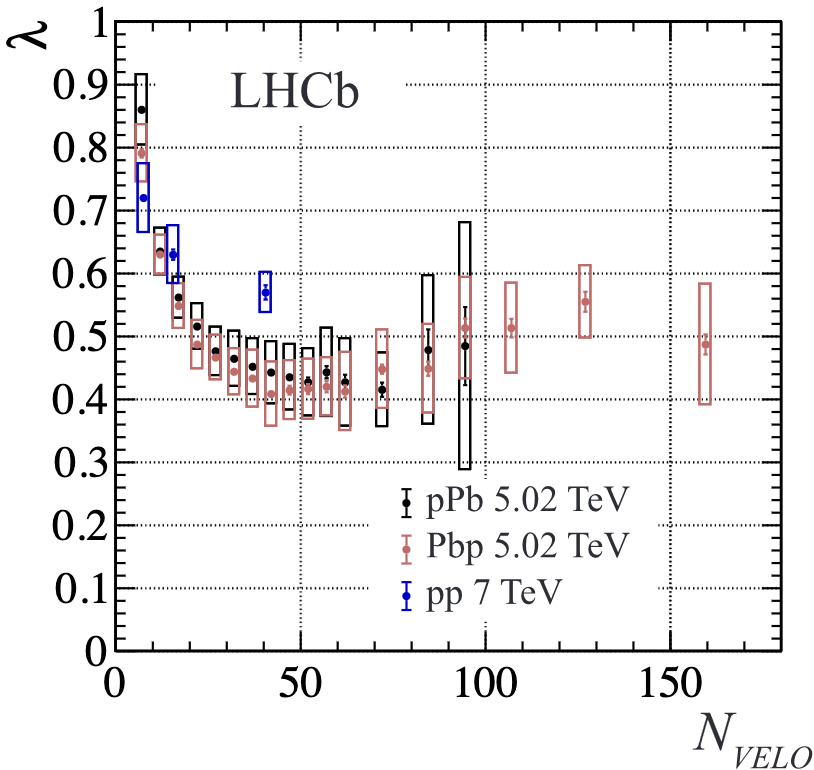}
    \caption{\label{fig:bec} Correlation radius $R$ (left) and strength $\lambda$ (right) as a function of the reconstructed charged-particle multiplicity $N_{\textnormal{\tiny VELO}}$ measured in the $pp$ and $p$Pb collision systems in the LHCb experiment. Error bars indicate the statistical uncertainties, while boxes illustrate the systematic ones. (left) Results of the fits to the observed radii linear in the cube root of the reconstructed multiplicity are indicated by the solid lines.}
\end{figure}

Multiparticle production within the process of hadronization has been investigated for six decades, but its nature is still not fully understood. In the case of joint production of identical bosons, the Bose-Einstein Correlations (BEC) arise and are observed in the form of
an enhancement of pair production for same-sign charged pions with small four-momentum difference squared $Q^2$. 
In particular, for small systems, such as those produced in proton-ion collisions, are interesting because the correlation lifetimes are significantly shorter than those in heavy-ion collisions, thus providing a better probe of the early system dynamics and the initial geometry of the hadron source. LHCb reports the first study of the BEC effect in $p$Pb collisions in the forward region~\cite{LHCb-PAPER-2023-002}. A two-particle correlation function is constructed as the ratio of the Q distributions for signal and the so-called reference pairs.
Signal pairs are formed from selected prompt, same-sign charged pions that originate from the same $pp$ collision vertex where the BEC are expected. The reference pairs are pairs of uncorrelated pions which reproduce as closely as possible the kinematics, Coulomb and non-femptoscopic effects present for the signal, except for the BEC effect. The correlation function is constructed in a format of ratio measurement, to cancel the effects related to the detector occupancy and material budget as well as to those related to the single-particle acceptance and efficiency. 
The correlation parameters are determined in bins of the charged particle multiplicity measured in the event. The measurements are performed in the full range of $Q\in [0.05,2.00] \;\mathrm{GeV}$. In general, the correlation radius of the spherical emitting source $R$ becomes larger with increasing event multiplicity, while the correlation strength $\lambda$ displays an opposite behaviour. The measured behaviour of the correlation parameters is compatible with observations from other experiments at LHC and is shown in \figurename~\ref{fig:bec}.
%
The measured correlation radii $R$ scale linearly with the cube root of the reconstructed charged-particle multiplicity and it is a tendency
compatible with predictions of hydrodynamic models on the system evolution~\cite{Bozek:2013df}.

\section{Charmonium and open charm production in $p$Ne collisions at $\sqrt{s_{NN}}=68.5\;\mathrm{GeV}$}

Charmonium suppression is a smoking gun for QGP (quark gluon plasma) production via the colour screening mechanism, which has been predicted more than 30 years ago~\cite{Matsui:1986dk}. This mechanism predicts the suppression of the $c\overline{c}$ bound states under the effect of QGP. Even though the $J/\psi$ suppression has been confirmed, the underlying mechanism cannot be completely understood due to cold nuclear matter effects (CNM) and statistical recombination effects at higher center of mass energies. The open charm production, such as $D^0$, serve as a reference and is a proxy for the overall $c\overline{c}$ pairs produced in the collisions. More measurements at different energies and with different colliding systems are fundamental to shade lights on the CNM, such as nPDFs~\cite{Eskola:2016oht} or comovers~\cite{Ferreiro:2012rq}. Additional measurements has been recently provided by LHCb~\cite{LHCb:2022tum,LHCb:2022lnf}, where the charmonium and $D^0$ production has been studied using the $p$Ne sample at $\sqrt{s_{NN}} = 68.5 \;\mathrm{GeV}$. 
The total integrated cross-section has been measured to be $\sigma_{J/\psi}^{y^* \in [-2.29,0]} = 506 \pm 8 \pm 25 \;\mathrm{nb}/\mathrm{nucleon}$, while $\sigma_{D^0}^{y^* \in [-2.29,0]} = 48.2 \pm 0.3 \pm 4.5 \;\mu \mathrm{b}/\mathrm{nucleon}$. This value has been extrapolated to the full phase space and found to be compatible with previous experiments. 
The HELAC-ONIA~\cite{Lansberg:2016deg} using CT14NLO and nCTEQ15 undershoot the data and good agreement is obtained with 1\% Intrinsic Charm (IC) contribution and without it~\cite{Vogt:2021vsc}, hence no conclusions on the presence or not of IC can be drawn here.  
The $D^0\textnormal{-}\overline{D}^0$ production asymmetry is
measured and it suggests a negative trend towards large negative $y^*$, where the valence quark contribution of the neon target is more significant.

\section{$J/\psi$ and $D^0$ production in PbNe collisions at $\sqrt{s_{NN}}=68.5 \;\mathrm{GeV}$}

\begin{figure}[htb]
    \centering
    \includegraphics[width=0.5\textwidth]{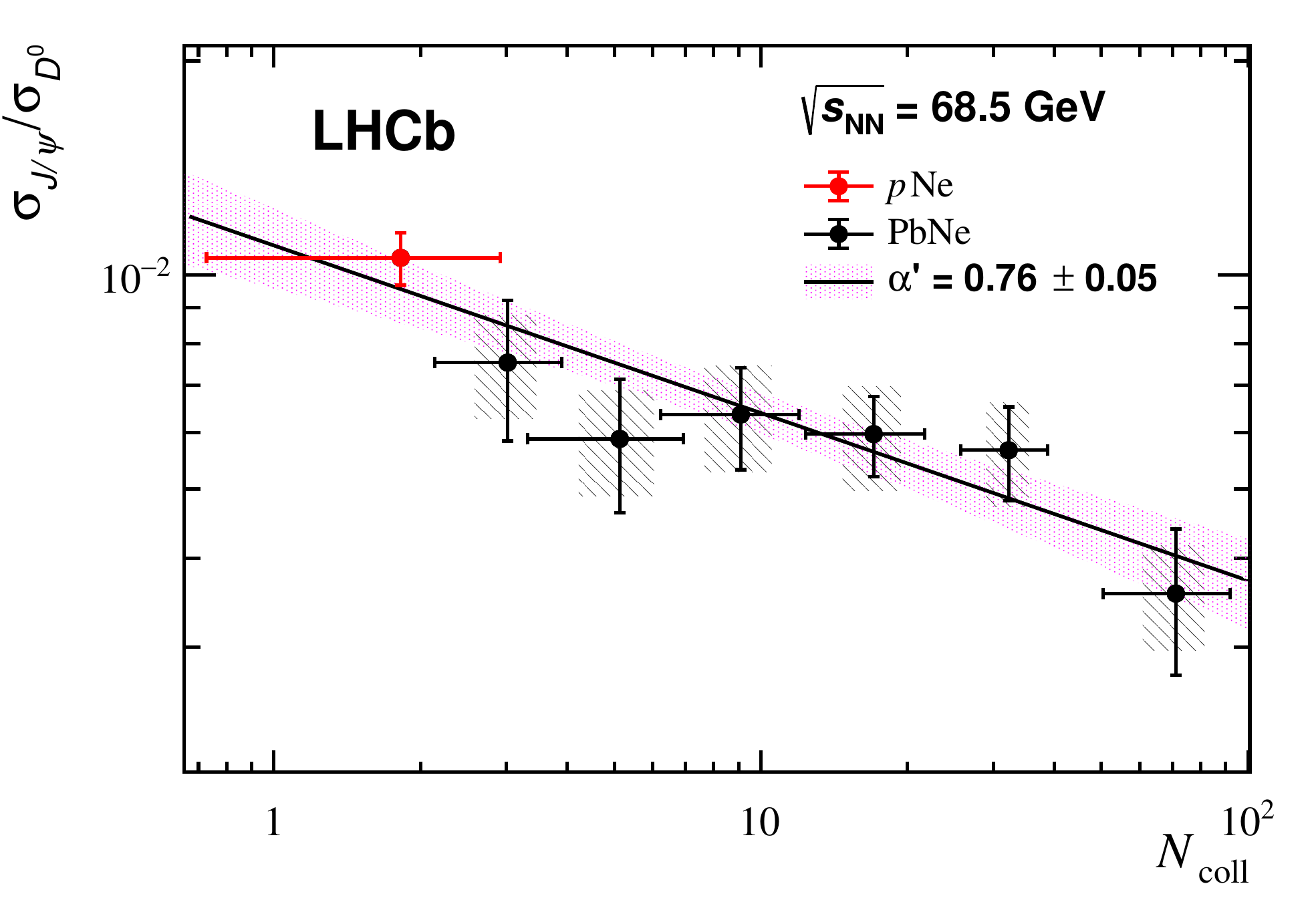}
    \caption{\label{fig:jpsi d0 production}$J/\psi$/$D^0$ cross-section ratio as a function of $N_{\textnormal{\scriptsize coll}}$; the error bars represent the quadratic sum of statistical and uncorrelated systematic uncertainties while the boxes correspond to the correlated systematic uncertainty. The red and black points correspond to $p$Ne and PbNe collisions, respectively.}
\end{figure}

The $J/\psi$ and $D^0$ cross-sections have been measured for the first time in fixed-target nucleus-nucleus collisions at a center of mass energy of $\sqrt{s_{NN}}=68.5\;\mathrm{GeV}$~\cite{LHCb:2022tyw}. The centrality in the sample has been determined using the energy deposit in the electromagnetic calorimeter. The $J/\psi$ has been reconstructed using the di-muon channel with approximately 545 signal events, whereas for the $D^0$ the two-body $D^0\to K^-\pi^+$ was used, with approximately 5670 signal events. The differential cross-section points out a strong dependence as a function of $p_T$. The integrated ratio of charmonium to meson is measured to be $\frac{\sigma_{J/\psi}}{\sigma_{D^0}} = (5.1 \pm 0.4 \pm 0.9) \times 10^{-3}$. The $\frac{\sigma_{J/\psi}}{\sigma_{D^0}}$ as a function of the number of binary nucleon-nucleon collisions $N_{\textnormal{\scriptsize coll}}$ (obtained from a Glauber model) has been evaluated and compared to measurements from $p$Ne collisions at the same center of mass energy in \figurename~\ref{fig:jpsi d0 production}.
The linear fit $\frac{\sigma_{J/\psi}}{\sigma_{D^0}} \propto N_{\textnormal{\scriptsize coll}}^{\alpha'-1}$ yields $\alpha' = 0.76 \pm 0.05$ which can be interpreted as a $J/\psi$ suppression consistent with CNM without additional QGP effects.

\section{Run 3 opportunities: the LHCb and SMOG upgrade}

The LHCb has undergone a major upgrade in 2022 to meet the challenge of the increased luminosity in the LHC Run 3~\cite{Bediaga:1443882}. The data acquisition is now based on a fully-software real-time event reconstruction and selection framework. The detector readout scheme is being upgraded to allow event processing at 40 MHz rate. The sub-detectors have been upgraded as well by introducing a new Vertex Locator (VELO) with pixel segmentation, new tracking stations and new RICH optics and PMTs. The upgrade will reduce the occupancy limitation in PbPb collisions and allow access to mid-central PbPb collisions up to 30\% in centrality. The other significant upgrade that will improve the heavy-ion physics programme is the new fixed- target storage cell located upstream of the LHCb interaction point, SMOG2. This new configuration allows injecting heavier and different noble gases (Kr, Xe, H2, D2, O2, N2) with a pressure about two orders of magnitude higher than the SMOG one, leading to higher luminosities for fixed-target collisions~\cite{LHCbCollaboration:2673690}. 
Since the two beam-beam and beam-gas interaction regions are well separated, now LHCb is the only experiment capable to operate simultaneously in collider and fixed-target modes at two energy scales.

\section{Conclusions and outlook}

The latest results obtained with the LHCb experiment in its fixed-target and heavy ion configuration have been discussed, including charmonium and open charm production in $p$Ne and PbNe fixed target collisions at $\sqrt{s_{NN}}=68.5 \;\mathrm{GeV}$ and BEC effects in $p$Pb collisions at $\sqrt{s_{NN}}=5.02 \;\mathrm{GeV}$. For the upcoming Run 3, LHCb will operate a new gas injection system called SMOG2 composed of a new storage cell allowing to locally increase the pressure of the target gas up to 35 times more than the gas pressure in SMOG. This will result in a significant increase in statistics of $\times 20$ ($\times 62$) more events with respect to the $p$He ($p$Ar) sample. The LHCb fixed-target physics program will be extended even further since SMOG2 allows to inject other species of gas, including heavier noble gases, as krypton and xenon, along with non-noble species such as hydrogen, deuterium or oxygen. 
Moreover, the new installed LHCb upgrade detectors will reduce the occupancy limitation in PbPb collisions and allow to access PbPb collisions up to 30\% in centrality.


\section*{Acknowledgments}

A.M. acknowledges support from the European Research Council (ERC) under the European
Union’s Horizon 2020 research and innovation programme under grant agreement No. 771642 (SELDOM) and INFN Sezione di Milano.

\section*{References}

\bibliography{refs}






\end{document}